\documentstyle[floats,aps,aipbook,epsf]{revtex}
\def\apj{{\it Ap.J.\rm }}
\def\apjl{{\it Ap.J. (Lett.)\rm }}

\def\apss{{\it Astr. Sp. Sci.\rm }}

{\catcode`\@=11                                                                 
\gdef\SchlangeUnter#1#2{\lower2pt\vbox{\baselineskip 0pt \lineskip0pt           
  \ialign{$\m@th#1\hfil##\hfil$\crcr#2\crcr\sim\crcr}}}}                        
\def\gtrsim{\mathrel{\mathpalette\SchlangeUnter>}}                              
\def\lesssim{\mathrel{\mathpalette\SchlangeUnter<}}

\def\erg{\varepsilon}

\def\teq#1{$\, #1\,$}                           

\def\vol#1#2{$\;$ \bf #1\rm , #2}
\def\pmb#1{\setbox0=\hbox{#1}%
  \kern-0.0125em\copy0\kern-\wd0
  \kern0.025em\copy0\kern-\wd0
  \kern-0.0125em\raise0.0433em\box0 }
\righthead{A.~K. Harding and M.~G. Baring}
\lefthead{Photon Splitting in Soft Gamma Repeaters}
\begin{document}
%
%
\title{A Photon Splitting Cascade Model of Soft Gamma-Ray Repeaters}
\author{Alice K.~Harding\footnote[1]{Email: \it Harding@lheavx.gsfc.nasa.gov}
   and Matthew G.~Baring\footnote[2]{Compton Fellow, Universities Space
   Research Association}}
\address{Laboratory for High Energy Astrophysics, Code 661,\\
   NASA Goddard Space Flight Center, Greenbelt, MD 20771, U.S.A.}
\maketitle
\begin{abstract}
The spectra of soft gamma-ray repeaters (SGRs), with the exception  of the
March 5, 1979 main burst, are characterized by high-energy cutoffs around 30 
keV and low-energy turnovers that are much steeper than a Wien spectrum.  
Baring\cite{Bar95} found that the spectra of cascades due to photon 
splitting in a very 
strong, homogeneous magnetic field can soften spectra and produce good fits 
to the soft  spectra
of SGRs.  Magnetic
field strengths somewhat above the QED critical field strength $B_{\rm cr}$, where $B_{\rm cr} = 4.413 \times 10^{13}$
G, is required to produce cutoffs at 30-40
keV. We have improved upon this model by computing Monte Carlo photon splitting
cascade spectra in a neutron star dipole magnetic field, including effects of
curved space-time in a Schwarzschild metric.  We investigate spectra produced
by photons emitted at different locations and observer angles.  We find that
the general results of Baring hold for surface emission throughout most of the
magnetosphere, but that emission in equatorial regions can best reproduce the
constancy of SGR spectra observed from different bursts.
\end{abstract}

\section{Introduction}

The association of supernova remnants with at least two of the three known
SGRs (2-4) 
(SGR1806-20 and SGR0525-66, the Mar 5, 1979 source) is now well 
established.  The third, SGR1900+14, is near a ROSAT source that is 
possibly a supernova remnant\cite{Hurley95}.  
These associations strongly indicate that
this class of $\gamma$-ray bursts is linked to relatively young galactic 
neutron stars.
Furthermore, it has been suggested\cite{DT92} that SGRs are
a special class of neutron stars, known as ``magnetars", that have
extremely strong magnetic fields.  Those neutron stars that are born with
periods of several ms can aquire high fields during a short period of vigorous 
convection that follows
core collapse.  The convection generates a rapid dynamo which can generate
dipole fields as high as $10^{15-16}$ G.  It has been pointed out that there 
are several attractive features of very high fields in accounting for 
observations of SGRs.  Among these are (i) an explanation of the 8 sec
periodicity of the Mar 5 event as dipole spin down in the $\sim 10^4$ yr
age of the N49 supernova remnant (ii) a reduction of the Compton scattering
opacity below the cyclotron fundamental, allowing a photon flux $\sim 10^4$
above the Eddington limit\cite{Pac92}.

Yet another advantage of very strong magnetic fields in models for SGRs
\cite{Bar95} is the effectiveness of photon splitting in producing the
soft spectra of SGRs.  Photon splitting, $\gamma 
\rightarrow \gamma\gamma$, attenuates $\gamma$-ray photons, degrading 
them to lower energies, where they split repeatedly until they escape
the high-field region.  Baring\cite{Bar95} showed that the emerging spectra
of such photon splitting cascades could account for both the shape and
the softness of observed SGR spectra if the field in the emission region
$B \gtrsim 4\,B_{\rm cr}$.  The splitting cascade spectra were computed 
by the solution of a kinetic equation for the photons, assuming a
uniform field in a region of size $R = 2 \times 10^6$ cm.  This approach,
while important in demonstrating the effectiveness of photon splitting
in modelling SGR spectra, neglected the dipole field geometry and the
strong gravitational field of a neutron star magnetosphere.  We have 
improved upon the model of Baring\cite{Bar95} by including
these effects in a Monte Carlo calculation of photon splitting cascades
near strongly magnetized neutron stars.  


\section{Photon Splitting Cascade Spectra}

Photon splitting is forbidden in field-free regions but is allowed
in neutron star magnetic fields.  The splitting rate\cite{Adler71} for
a photon of energy $\epsilon$ in units of $mc^2$,
$T_{sp} (\erg ) \propto \erg^5\,(B/B_{\rm cr})^6\,\sin^6\theta_{\rm kB}$,
where $\theta_{\rm kB}$ is the angle between the photon momentum and the
magnetic field, can be large if $B$ is near $B_{\rm cr}$.
The above rate is valid in the nondispersive limit, where 
\teq{\erg\sin\theta_{\rm kB} \lesssim 2} and \teq{B\ll B_{\rm cr}}.
It is dependent on the two polarization states of the photons in the 
birefringent, magnetized vacuum: $\parallel$ or $\perp$, where the photon's
electric vector is parallel or perpendicular to ${\bf k \times B}$. 
The modes $\parallel \rightarrow \perp
\parallel$, $\perp \rightarrow \perp\perp$ and $\perp \rightarrow \parallel
\parallel$ are the only ones that are non-zero through CP invariance. 
Since there is a difference in the rates of these modes which depends on field
strength\cite{Bar95}, polarized photons emerge from the emission region. 
In our
calculations, we have included high field corrections to the above formula
for splitting\cite{Adler71} that cause the attenuation
coefficient to saturate somewhat above $B \sim 4 B_{\rm cr}$.  

\begin{figure} 
\centering \leavevmode 
\hbox{\hbox{\epsfxsize=0.75\columnwidth \epsfbox{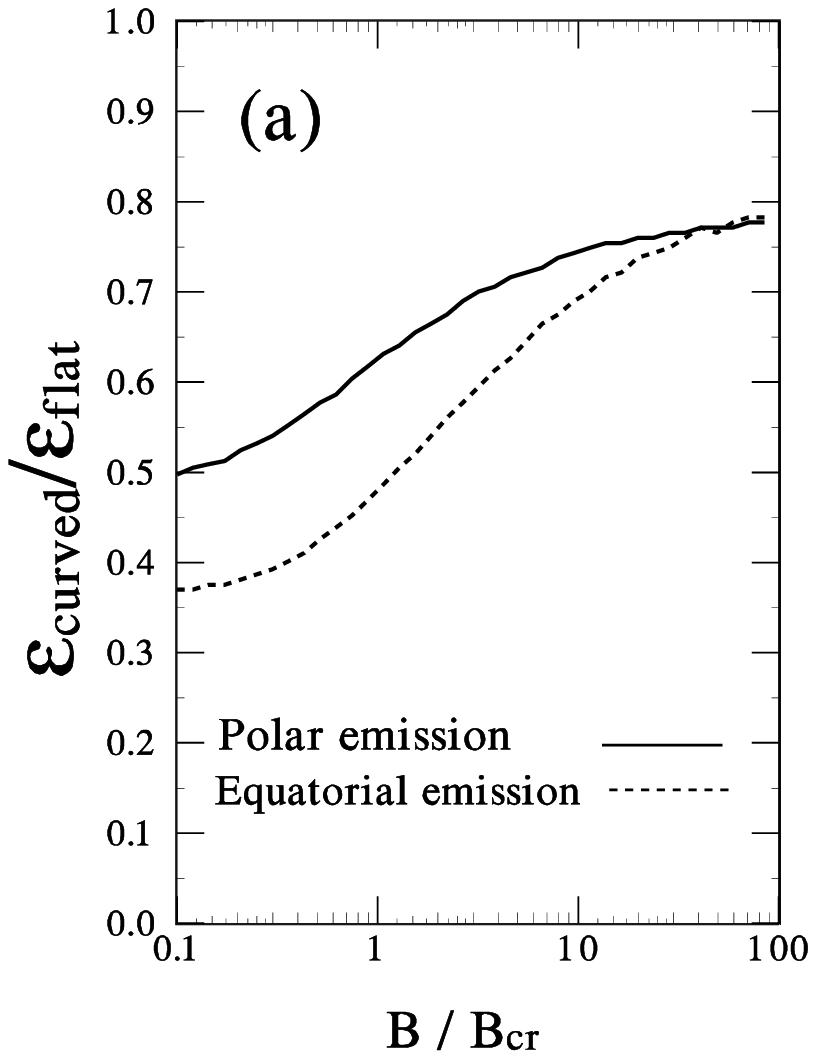}}
\hfil\hskip -1.7in
\hbox{\epsfxsize=0.7\columnwidth \epsfbox{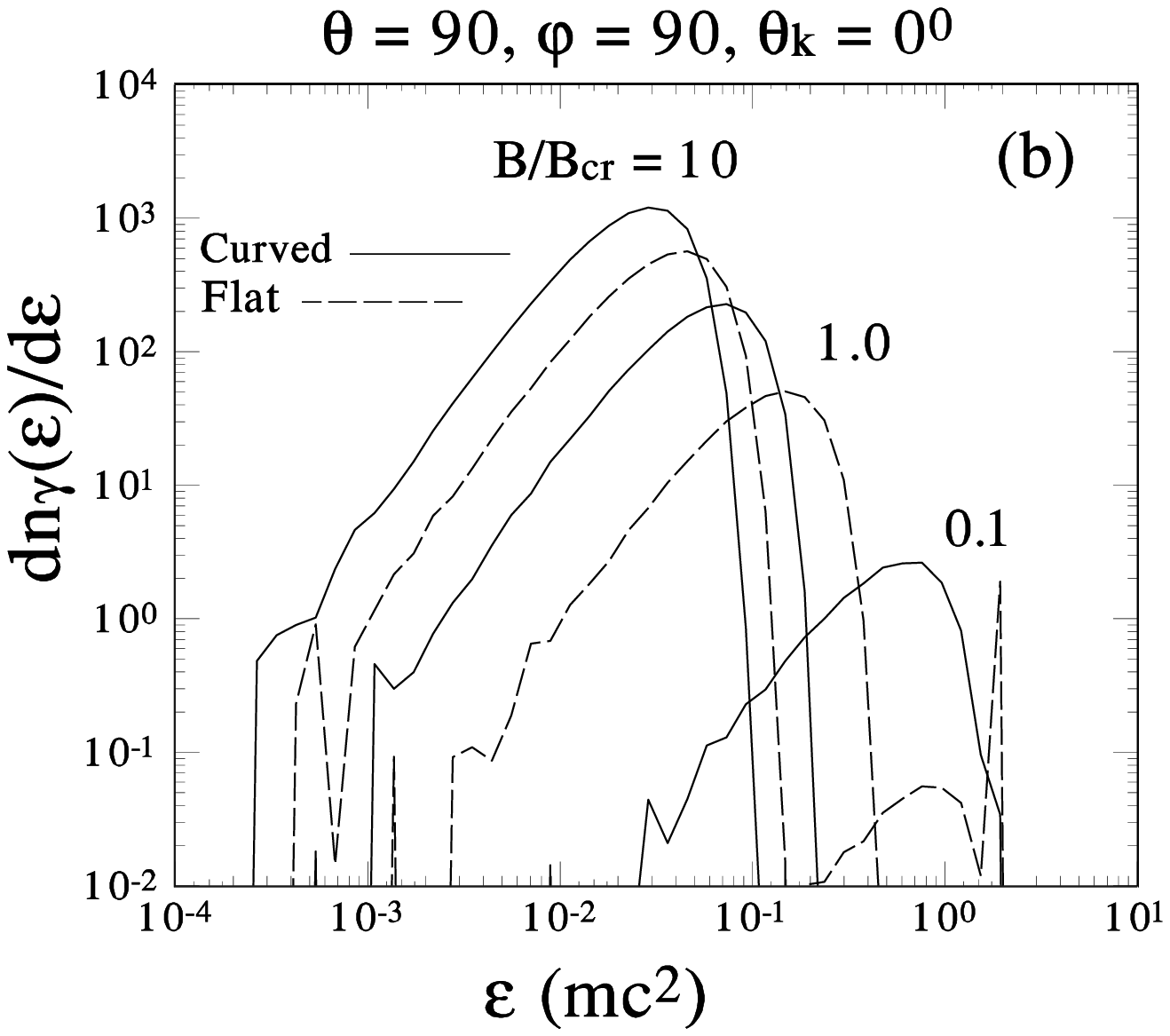}}}
\vskip 0.2truecm
\caption{a) Ratio of the energies below which photons escape a dipole
field without splitting in curved to flat space as a function of magnetic field. b) Photon splitting cascade spectra in curved and flat space for
different surface field strengths.  $\theta$ and $\phi$ are the 
colatitude and azimuth angles of 
the photon emission point in the dipole field and $\theta_{\rm k}$ is the
photon emission angle to the dipole axis in the local inertial frame.}
\end{figure}

Photon splitting cascade spectra will turn over roughly at the 
escape energies of the photons, above which photons undergo at
least one splitting generation, but below which the optical depth is
always $\ll 1$ and photons can escape the magnetosphere.  
The existence of such an escape energy is a consequence
of the $r^{-3}$ decay of the dipole field. 
Our previous analysis\cite{BH95} showed that escape energies
for photons emitted at the neutron star surface and propagating through
a dipole magnetic field are quite sensitive to the propagation angle 
$\theta_{\rm k}$ of 
the photon to the dipole axis at the magnetic pole 
($\theta = 0^0$),
but nearly independent of this angle near the equator ($\theta = 90^0$).
This effect is a function of the dipole field curvature:
near the pole, the field lines are diverging rapidly, and different emission 
directions sample very different field orientations, but at the equator
the field looks nearly the same in all directions.   

\newpage
In the present work we have included the general relativistic effects of
curved spacetime in a Schwarzschild metric, following the treatment of
Gonthier \& Harding\cite{GH94} who studied the effects of general relativity
on photon attenuation via magnetic pair production.   Those effects are 
the curved
spacetime photon trajectories, the magnetic dipole field in a Schwarzschild 
metric and the gravitational redshift of the photon energy as a function of
distance from the neutron star.  We have taken a neutron star mass,
$M = 1.4\,M_\odot$ and radius, $R = 10^6$ cm in these calculations.
The effects of curved spacetime decrease the escape energies by a factor
of about 2 compared to flat spacetime\cite{BH95}, 
the largest contributions coming
from an increase in the dipole field strength (by $1.4$ at the pole) and
the correction for the gravitational redshift, which increases the photon 
energy by roughly a factor of 1.2 in the local inertial frame at the
neutron star surface.    
The qualitative behavior of escape energies as a function of 
$\theta_{\rm k}$ in curved space is the same as in flat space: 
in the polar emission 
case, there is still a substantial
variation in escape energy with emission direction $\theta_{\rm k}$
for both initial polarizations, but almost no variation in the case
of equatorial emission.  The ratio of curved to flat space escape energies
as a function of magnetic field are shown in Figure 1a.  At low field 
strengths, one is seeing the combined effects of the increase in the
dipole field strength (which varies from pole to equator) and the increase 
in local frame photon energy; in high fields, only the increase in photon
energy affects the escape energy, giving a factor of 1.2 decrease, due
to the saturation of the splitting with increasing field strength.  


Figure 1b shows Monte Carlo cascade spectra resulting from monoenergetic
photons injected with $\epsilon = 2$ at the neutron star surface at the 
magnetic pole.  The
photons split many times before escape, each time dividing their energy into
two photons with a distribution that peaks
at half the parent photon energy, in the non-dispersive limit (8,9).  
The cascade
spectra peak just below the escape energy for that field strength.  
Compared to those of Baring, these spectra show the same $\epsilon^2$
power law below the peak, but a more rapid decrease 
above the peak than the $\epsilon^{-7}$ found by Baring, due to the fact that
the Monte Carlo spatial injection occurs as a delta function at the surface, 
while the kinetic equation solution assumes an exponential injection.
The peak in the flat space spectra in Fig. 1 are also a factor of around 
2 higher than the homogeneous field case of Baring, due to the fall off in
strength of the dipole field.  The cascade saturates at $B \sim 30 B_{\rm cr}$,
due to the saturation of the cross section in high fields, but the saturation
energy as well as the peaks of all spectra are lower when curved 
space corrections are included.

\begin{figure} 
\centering \leavevmode 
\hbox{\hskip -0.4truecm
\hbox{\epsfxsize=0.7\columnwidth \epsfbox{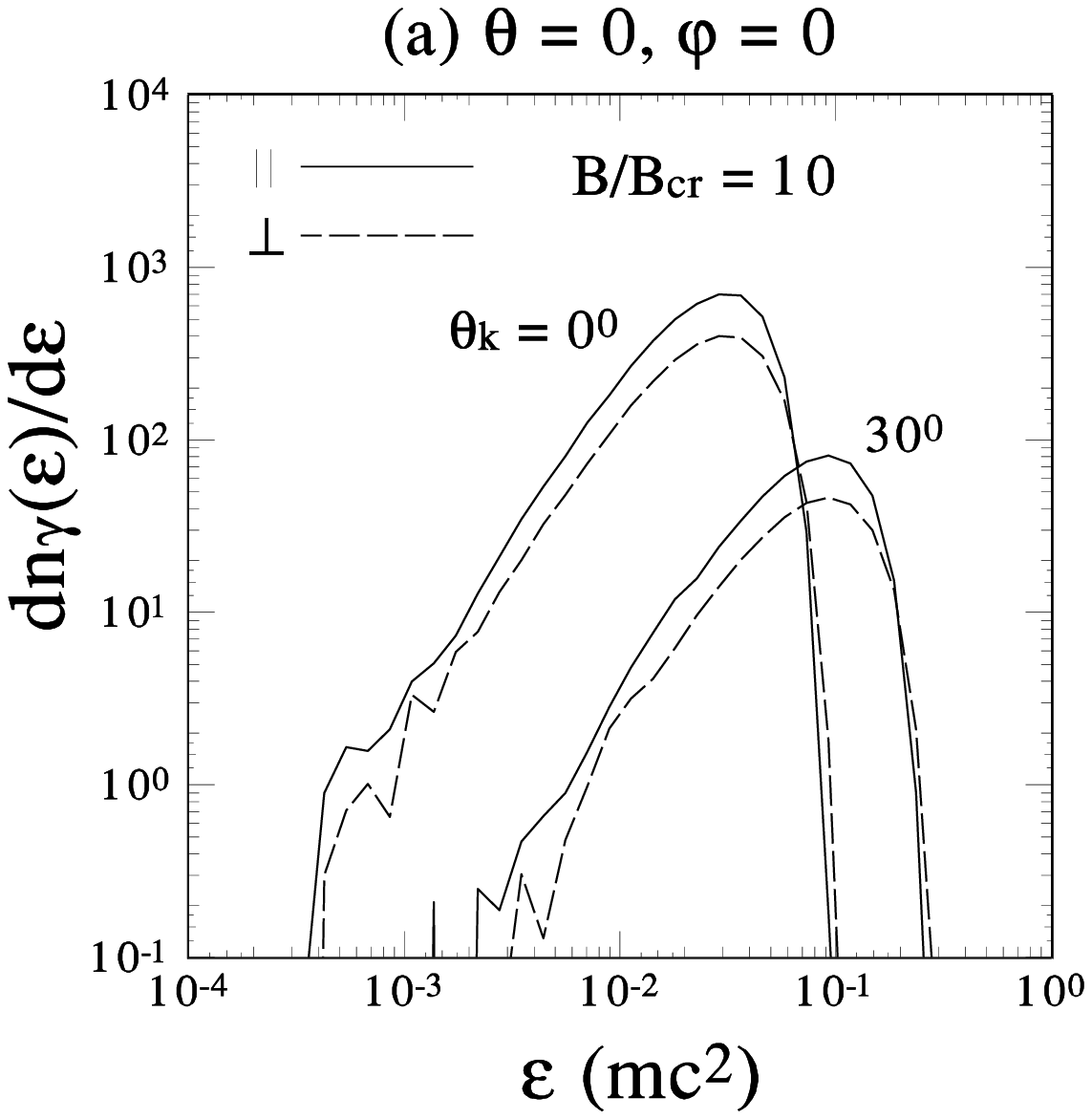}} \hfil
\hskip -0.9in
\hbox{\epsfxsize=0.7\columnwidth \epsfbox{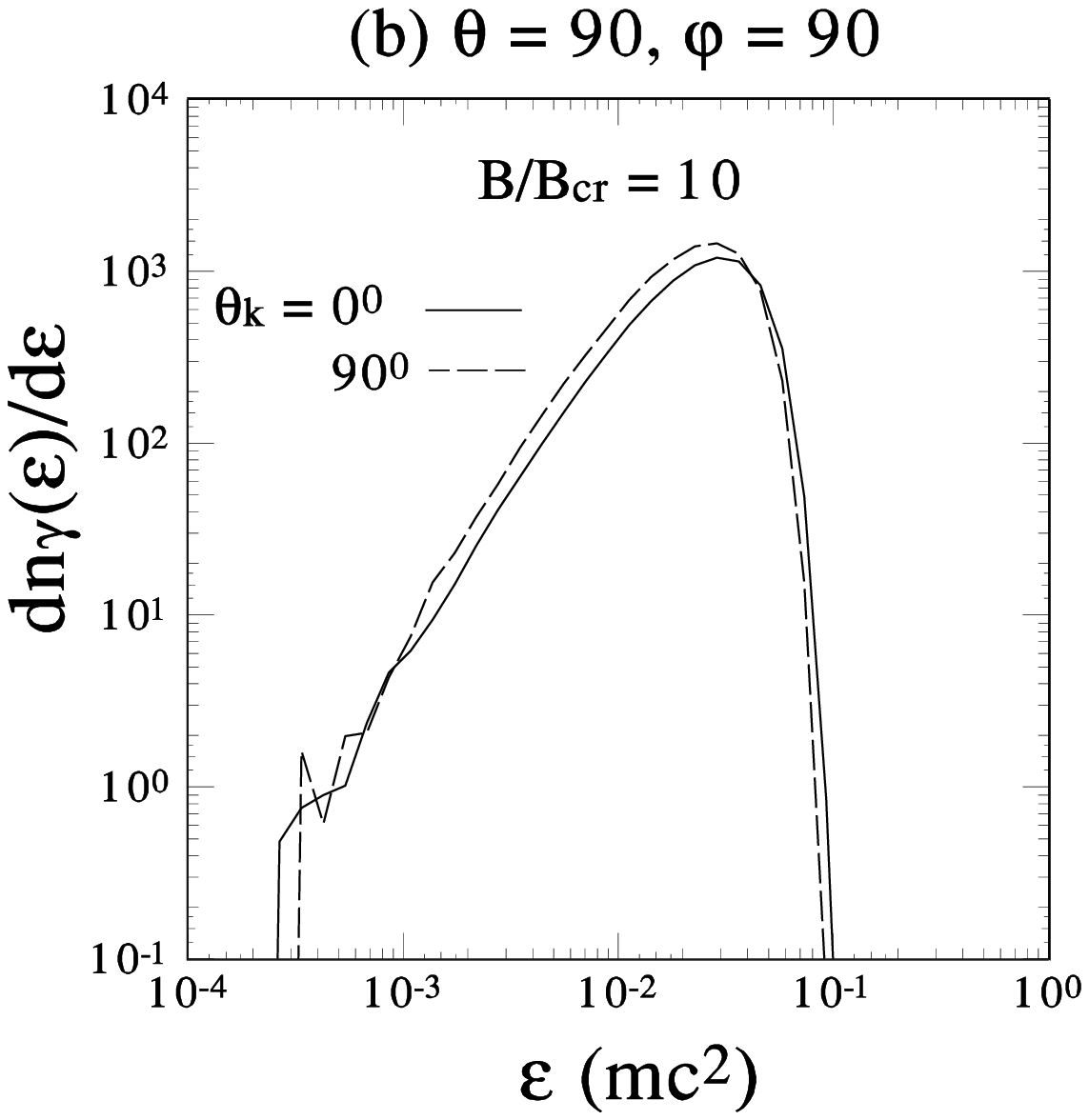}}}
\vskip 0.2truecm
\caption{Photon splitting cascade spectra in curved space for photons 
emitted at the (a) pole and 
(b) equator of a dipole field and for different emission directions 
$\theta_{\rm k}$ to the dipole axis.}
\vspace{-0.2cm}
\end{figure}

The effect of varying the emission location on the neutron star surface on
the cascade spectrum is shown in Figure 2.  Fig. 2a illustrates that the
spectra are distinctly different for two emission angles at the magnetic pole,
while they are almost identical for different observer angles at the
equator.  Thus, the behavior of the cascade spectrum is almost entirely
determined by the escape energies at the initial photon injection point.

\section{Discussion}

The generation of \it pure \rm splitting cascades is contingent upon operation
of at least two polarization modes of splitting so that polarization exchange
is effected.  Adler \cite{Adler71} demonstrated that in the dispersive
magnetized vacuum, only one of the modes of splitting considered here,
namely $\perp\to\parallel\parallel$, satisfies kinematic selection rules 
imposed by
four-momentum conservation.  This result was derived assuming weak dispersion,
i.e. a refractive index close to unity.  While Shabad \cite{Shabad75} has
extensively looked at the regime of strong vacuum dispersion, mostly near and
above pair creation threshold, numerical computation of the refractive
index and kinematic selection rules for supercritical fields well below pair
threshold remains to be explored; this is a major goal of our future research.

Should the selection rules extend to the strongly dispersive regime sampled
by the SGR problem, cascade continuation could be effected by a
polarization-switching process such as Compton scattering.  The cross-section
for this is suppressed below the cyclotron resonance \cite{Herold79}, and
effective polarization state switching ($\parallel \to \perp$) will occur 
if the radiation is somewhat
beamed along the field in the rest frame of the scattering electrons.
Significant scattering opacities are quite plausible in the luminous 
environment of SGRs, and the complications of scattering in a dipole field
geometry, which will tend to broaden the ouput cascade spectrum and smear
its polarization spectrum, are deferred to future work.

These calculations of photon splitting cascades in 
neutron star magnetospheres lend support to the idea that the splitting
mechanism in strong magnetic fields could be the cause for the softness
and quasi-thermal shape of SGR spectra.  The insensitivity of the
cascade spectra to observer angle near the equator is very important
for modeling SGR spectra.  It implies that the observed spectra would not
vary from burst to burst, even if the neutron star orientation changes
(i.e. the star rotates), as long as the emission occurs at large
magnetic colatitudes.  This is consistent with the lack of observed
spectral variation in bursts from SGR1806-20\cite{FLM94} and in the 
phase-resolved spectroscopy of the periodic soft
tail of the Mar 5, 1979 event\cite{Mazets82}.


\end{document}